\documentclass[referee]{raa}
\usepackage{graphicx,times}
\usepackage{natbib}
\usepackage{amssymb,amsmath}
\bibpunct{(}{)}{;}{a}{}{,}

\usepackage[a4paper=true,dvipdfm=true,pagebackref=true]{hyperref}
\hypersetup{pdftitle = The title of my PDF, pdfauthor = My name, pdfsubject= The subject, pdfkeywords = keyword1 keyword2 keyword3}
\hypersetup{colorlinks = true, linkcolor = green, anchorcolor = red, citecolor = blue, filecolor = red, pagecolor = red, urlcolor = red}

\begin{document}

   \title{Can We Determine the Filament Chirality by the Filament Footpoint Location or the Barb-bearing?}

%\footnotetext{\small $*$ Supported by the National Natural Science Foundation of China.}

 \volnopage{ {\bf 2015} Vol.\ {\bf X} No. {\bf XX}, 000--000}
   \setcounter{page}{1}

   \author{Q. Hao\inst{1,2,3,4}, Y. Guo\inst{1,2}, C. Fang \inst{1,2}, P. F. Chen\inst{1,2}, W. Cao\inst{3,4}
}
   \institute{ School of Astronomy and Space Science, Nanjing University, Nanjing 210093, China; {\it fangc@nju.edu.cn}\\
        \and
             Key Laboratory for Modern Astronomy and Astrophysics (Nanjing University), Ministry of Education, Nanjing 210093, China\\
	\and
	New Jersey Institute of Technology, Center for Solar Research, 323 Martin Luther King Blvd., Newark, NJ, U.S.A.\\
\and
Big Bear Solar Observatory, 40386 North Shore Lane, Big Bear City, CA, U.S.A.\\
\vs \no
   {\small Received 2015 xx xx; accepted 2015 xx xx}
}

\abstract{We attempt to propose a method for automatically detecting the solar filament chirality and barb bearing. We first introduce the unweighted undirected graph concept and adopt the Dijkstra shortest-path algorithm to recognize the filament spine. Then, we use the polarity inversion line (PIL) shift method for measuring the polarities on both sides of the filament, and employ the connected components labeling method to identify the barbs and calculate the angle between each barb and the spine to determine the bearing of the barbs, i.e., left or right. We test the automatic detection method with H$\alpha$ filtergrams from the Big Bear Solar Observatory (BBSO) H$\alpha$ archive and magnetograms observed with the Helioseismic and Magnetic Imager (HMI) on board the \textit{Solar Dynamics Observatory} (\textit{SDO}). Four filaments are automatically detected and illustrated to show the results. The barbs in different parts of a filament may have opposite bearings. The filaments in the southern hemisphere (northern hemisphere) mainly have left-bearing (right-bearing) barbs and positive (negative) magnetic helicity, respectively. The tested results demonstrate that our method is efficient and effective in detecting the bearing of filament barbs. It is demonstrated that the conventionally believed one-to-one correspondence between filament chirality and barb bearing is not valid. The correct detection of the filament axis chirality should be done by combining both imaging morphology and magnetic field observations.
\keywords{Sun: filaments, prominences --- Sun: magnetic fields  --- Sun: chromosphere --- techniques: image processing
}
}

   \authorrunning{Qi Hao et al. }            %author_head in even pages
   \titlerunning{Determine the filament chirality}  % title_head in odd pages
   \maketitle

%________________________________________________ sections below
%
\section{Introduction}
Solar filaments, called prominences when they appear above the solar limb, are plasma in the low corona that is one hundred times cooler and denser than its surroundings.They are particularly visible in H$\alpha$ observations, where they often appear as elongated dark features with several barbs or feet \citep{1995Tandberg}. Filament barbs can be traced to the legs of prominences when observed above the limbs against the solar disk. They have an average asymmetry with respect to the long axis of a filament, known as the ``spine"; the spine defines the main body of a filament and runs its full length. Filament barbs tend to be broader near the filament axis and converge to chromospheric points downward to the chromosphere. Martin and her colleagues \citep{1994Martin,1998Martin} gave reasonable classifications for the filament spine chirality and barb bearing: the filament whose axial field is directed rightward when viewed by an observer at the positive-polarity side is called dextral, and the one whose axial field is directed leftward from the same perspective is called sinistral; when viewed from either footpoint of a filament, if the barbs on the near one veer from the filament spine to the right (left), the barbs are defined as right-bearing (left-bearing). Dextral and sinistral filament axis chirality correspond to negative and positive magnetic helicity, respectively. Therefore, the filament axis chirality tells directly the magnetic helicity.

Filaments are always aligned with photospheric magnetic polarity inversion lines (PIL) and magnetic flux cancellation also occurs at the photosphere close to it \citep{1998Martin}. Filaments sometimes undergo large-scale instabilities, which break their equilibria and lead to eruptions; therefore, they are often associated with flares, coronal mass ejections (CMEs) and other solar activities \citep{2000Gilbert,2003Gopalswamy,2004Jing,2008Chen,2011Chen,2011Shibata,2011Green,2012Webb,2012Zhang,2014Zhou,2014Ma}. Hence, understanding the magnetic configuration of solar filaments is important in resolving the relation of filaments to flares and CMEs as well as the formation of filaments themselves. In traditional theoretics, filament material is supported against gravity by Lorentz force in magnetic dips, which are formed by either dipped arcade loops \citep{1957Kippenhahn} or twisted magnetic flux rope \citep{1974Kuperus}.  \citet{1994Martin} and \citet{1998Martin} constructed a wire model. In their model, barbs are field line threads which are rooted in the minor polarity flux concentrations in the photosphere and a filament consists of many threads that are highly sheared. \citet{1998Aulanier} and \citet{1998Aulanier2,1999Aulanier} constructed a series of models based on the linear force-free field and magnetohydrostatic extrapolation of the observed line-of-sight magnetic fields in the photosphere. On the contrary, a three-dimensional flux rope model is able to reproduce the observed filament barbs and the barb bearing patterns with parasitic polarities \citep{2000Aulanier}. In these models the barbs are interpreted as dips in the flux rope resulting from the interaction of the low-lying portion of the transversal fields in the twisted flux rope with photospheric parasitic polarities on either side of the filament channel. These models imply that the filament barbs are observed morphological features that can be used to estimate magnetic helicity\citep{1998Martin}. However, filaments with both left-bearing and right-bearing chirality barbs along the their axes have been observed \citep{2003Pevtsov,2010Guo1,2010Chandra}, which means we cannot determine the magnetic helicity only by the morphological features in an H$\alpha$ filtergram.  The magnetograms revealing the magnetic fields in and around the filament channel should also be taken into consideration.

Developing and using appropriate methods for automatically detecting the characteristics of filaments is imperative since the time cadence and spatial resolution of the observations are becoming higher and higher and there are a huge amount of data that cannot be analyzed by human visualization. Since \citet{2002Gao} combined the intensity threshold and region growing methods in order to develop an algorithm to automatically detect the growth and the disappearance of filaments, a number of automated filament detection methods and algorithms have been developed in the past decade \citep{2003Shih,2005Fuller,2005Bernasconi,2005Qu,2010Wang,2010Labrosse,2011Yuan,2013Hao}, and these algorithms can be combined with image enhancement techniques \citep[e.g.,][]{feng14}. While for more specific observation features, such as the barbs and the spines of filaments, effective automatic detection methods are few. \citet{2005Bernasconi} developed a method applying a principal curve algorithm based on the geometric theory, which was recently updated by \citet{2012Martens}, to determine the filament spine, then to calculate the distance between the boundary and the filament spine in order to find barbs. They considered nearly one year's observation of filaments and the results confirm the hemispheric magnetic helicity rule. \citet{2011Yuan} designed a method to find the filament spines based on the graph theory which is also used in our spine detection method.

In this paper, we try to present an efficient automated method for detecting solar filament spine and the barb bearings. In Section~\ref{methods}, we describe the details of our automatic detection algorithm for the solar filament spine, barbs, and the corresponding chirality. We give four examples in Section~\ref{results}. We present our discussions and draw our conclusions in Section~\ref{conclusion}.

\section{Methods}
\label{methods}
In our previous work \citep{2013Hao}, we developed an efficient and versatile automated detecting and tracing method for solar filaments. Here, we still use this method to detect filaments in H$\alpha$ full disk filtergrams, then we extract each filament for further processing by the newly developed method. The details are explained in the following subsections.

\subsection{Filament Axis Chirality Detection}
\label{axis}

\subsubsection{Filament Spine Detection}
\label{spine}

\begin{figure}
\centering
\includegraphics[width=0.4\textwidth,clip=]{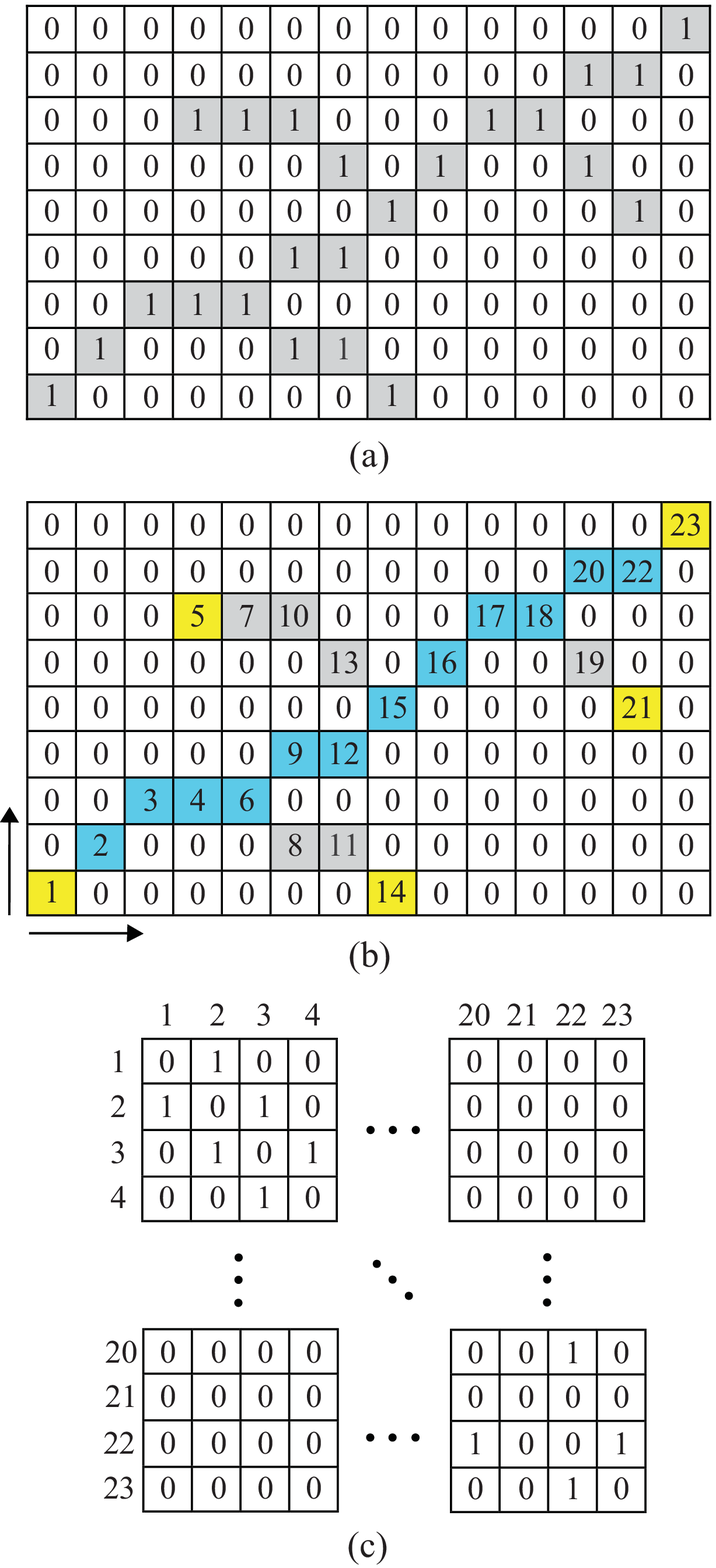}
\caption{ An example for explaining the filament spine automatic detection algorithm. (a) Filament skeleton nodes are the pixels whose values are equal to 1 (the values of the background pixels are equal 0). In order to facilitate the distinction, the filament skeleton nodes are dyed gray. (b) Each skeleton node is marked by a unique index number, from 1 to 23. The yellow nodes are the vertex nodes. The nodes with blue color, including two yellow vertex nodes No. 1 and No. 23 constitute the longest path among the every 2 of the 5 vertex nodes which is determined as the filament spine. The arrows shows the algorithm traversal sequence that start at the lower left corner from bottom to top and from left to right. (c) The four corner parts of the whole connection matrix. Because the size is $23 \times 23$, which is too large to show, only the four corners are shown. All the detail steps are described in the text.
} \label{fig1}
\end{figure}

After processing the H$\alpha$ full disk images, we obtain each filament skeleton binary image. Then, we use the morphological spur removal method, which employs the morphological hit-and-miss transformation ( a general binary morphological operation which is useful in locating particular pixel configurations in an image. It takes as input a binary image and structuring element pairs, and produces another binary image as output. The details of this method can be found in Section 4.4 of \citet{2013Hao}), to find and remove the barbs for several iterations to obtain the filament spine. In order to get a thin filament spine without any barbs, the steps of iterations need to be sufficient. However, more iterations would bring a disadvantage that the filament spine may be short than it really is. To avoid this error, we adopt a new method using the graph theory as done by \citet{2011Yuan}. This method regards the filament skeleton as an unweighted undirected graph, which contains nodes (i.e., pixels in the filament skeleton) and connectivity of the nodes. If two nodes of the skeleton are 8-connected (considering pixels in a $3\times3$ matrix, if current pixel is \textit{(i,j)} which is in the middle of the matrix, then the pixels around it, i.e., \textit{(i-1,j-1)}, \textit{(i-1,j)}, \textit{(i-1,j+1)}, \textit{(i,j-1)}, \textit{(i,j+1)}, \textit{(i+1,j-1)}, \textit{(i+1,j)}, \textit{(i+1,j+1)}, are 8-connected with current pixel \textit{(i,j)}), it means that there is an edge between them. The longest acyclic path of two nodes is defined as the filament spine. One can find the spine without changing the topological structure of the filament skeleton. The main algorithm is described as following. The algorithm is pictorially shown by a sample filament skeleton example in Figure~\ref{fig1}.

Firstly, for each binary image with a single filament skeleton, we have to find all the nodes (i.e., pixels) of the filament skeleton and assign index numbers to them. Since a real observed filament is relatively large and complex, in order to explain our method clearly, here we take an artificial filament skeleton as an example, which has the same typical features as real ones. As the model example shown in Figure~\ref{fig1}(a), filament skeleton nodes are the pixels whose values are equal to 1 (the values of the background pixels are equal to 0). Figure~\ref{fig1}(b) shows that each filament skeleton node is assigned with a unique index number. This example has 23 nodes in total. The algorithm goes through each pixel starting at the lower left corner from bottom to top and from left to right, as indicated by the arrows shown in Figure~\ref{fig1}(b).

Secondly, we build an $N \times N$ connection matrix that records the edge information between each nodes, where $N$ is the number of the filament skeleton nodes. If the $i$th node and  the $j$th node are 8-connected, the values of the corresponding coordinates $(i,j)$ and $(j,i)$ in connection matrix are 1. It means that there is an edge between them, namely, a connection exists between the two nodes. Otherwise the value is 0, which indicates that the two nodes have no edge between them. In addition, self connecting nodes are not allowed here. Therefore, we ignore the diagonal of the connection matrix and set these values as 0 for convenience. Next, for each node of the filament skeleton, we check its connectivity with other nodes and update the corresponding values in the connection matrix. Figure~\ref{fig1}(c) shows only four corner parts of the whole connection matrix since the size of the whole matrix is $23 \times 23$, which is too large to be shown completely. For instance, we can see that the node No. 2 just connects with nodes Nos. 1 and 3 in Figure~\ref{fig1}(b), so the corresponding values of the coordinates $(1,2)$, $(2,1)$, $(2,3)$ and $(3,2)$ in connection matrix are 1. Meanwhile, the other coordinates in the second row and the second column are 0, which means that these nodes are not connected with node No. 2.

Thirdly, we have to find the vertex nodes (i.e., pixels in the ends of the filament skeleton). Vertex nodes are defined as those nodes that only have one connected neighbour, namely, each of them has only one edge associated with it. The vertex nodes are easy to be found in the connection matrix since the corresponding row or column has only one item whose value is equal to 1. The example in Figure~\ref{fig1}(b) shows the vertex nodes in yellow pixels, i.e., nodes Nos. 1, 5, 14, 21, and 23.

Fourthly, we use the shortest-path method of the graph theory to find all the paths from one vertex node to other vertex nodes. The longest acyclic path is the spine. There are 5 vertex nodes so that the program finds the shortest-path between every 2 of the 5 vertex nodes and obtains 10 paths. Among them, the longest path has 13 edges that are from node No. 1 to node No. 23. This path is determined as the spine of the example filament as shown in Figure~\ref{fig1}(b) with  blue color, including the two vertex nodes No. 1 and No. 23 that are treated as the footpoints of the filament.

%figure 1

\begin{figure}
\centering
\includegraphics[width=0.4\textwidth,clip=]{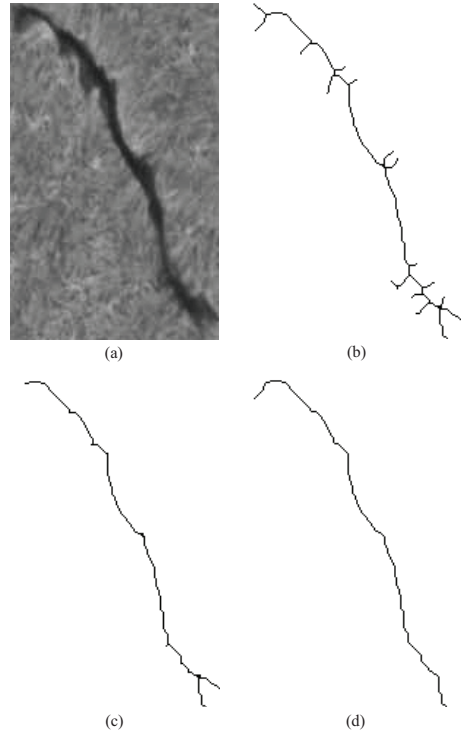}
\caption{ An example showing the determined spine. (a) A filament image extracted from the enhanced H$\alpha$ full disk filtergram. (b) The skeleton image. (c) The filament spine determined by the morphological spur remove method. (d) The filament spine determined by the shortest-path method.  }
\label{fig2}
\end{figure}

The shortest-path method determines the single-source shortest path from one node to all other nodes in the graph represented by the $N \times N$ connection matrix, where $N$ is the number of nodes. It can get the spine as thin as one pixel. Here, we use the Dijkstra algorithm \citep{1959Dijkstra} to find the shortest paths. The time complexity is $O(\log (N) \ast E)$, where $E$ is the number of edges. Note that the connection matrix is sparse, which can be compressed to reduce the storage space. An example is shown in Figure~\ref{fig2}. Figure~\ref{fig2}(a) shows a filament image extracted from the enhanced H$\alpha$ full disk filtergram and Figure~\ref{fig2}(b) shows its skeleton image. For comparison, we first adopt the morphological spur removal method and process the morphological hit-and-miss transformation to find and remove the barbs for 10 iterations. The filament spine is then obtained, as shown in Figure~\ref{fig2}(c). We find that the spine is shorter than the original filament and small spurs still remain. Especially, a barb still remains in the right bottom corner. Alternatively, Figure~\ref{fig2}(d) shows the spine determined by the shortest-path method of the graph theory. Compared with the filament skeleton in Figure~\ref{fig2}(c), the spine in Figure~\ref{fig2}(d) is much thinner and runs the full length of the filament without changing the topological structure of the filament skeleton.

%figure 2

\subsubsection{Filament Axis Chirality}
\label{axis chirality}
The filament axis chirality is defined as that, when viewed from the positive polarity of the magnetic field, if the axial magnetic field of a filament points to the left (right),  the filament is sinistral (dextral). It can be determined by measuring the magnetic field along the axis of a filament relative to the main polarities \citep{1998Martin}, which in principle requires the vector magnetograms and even the nonlinear force-free field extrapolation. In practice, \citet{1994Martin} proposed that the chirality of a filament can be obtained by measuring the magnetic polarities of the two footpoints of the filament. When doing so, we should acquire four basic magnetic field measurements, i.e., the magnetic polarities of the two footpoints of the filament and the dominant magnetic polarities along both sides of the filament spine. The polarities of the two footpoints can be easily determined by the direct measurements of the magnetic field once the positions of the footpoints of the H$\alpha$ filament are recognized automatically, while the polarities on both sides of the filament are relatively more difficult to be measured and the measurements have many uncertainties due to the complexity of the magnetic field. Here, we propose a PIL shift method to derive the magnetic polarities on both sides of the filament spine  relatively more precisely. The procedure of the algorithm is detailed as below.

Firstly, we co-align the observed magnetogram with the H$\alpha$ filtergram. The magnetic polarities of the two filament footpoints are measured at the coordinates of the filament footpoints, which were already derived by the automatic spine detection method from the H$\alpha$ filtergram. The two footpoints are designated by P and N, respectively, meaning positive and negative polarities.

Secondly, the PIL coordinates along the filament spine are calculated and obtained. In order to get a single and thin PIL, we adopt proper smoothing of the original magnetogram. In our process, the smoothing window is around 15 pixels wide. Then, we set the positive magnetic footpoint as the coordinate origin, and then calculate the slope of the straight line connecting the two footpoints of the filament, i.e., line PN. The PIL coordinates are then shifted sideward along the direction perpendicular to line PN with a given distance $s$. In our following process, we set $s$ to be $10\%$ of the spine length. The magnetic field polarities on both sides are measured along the shifted PIL points.

Finally, we check the magnetic polarities on the two sides of line PN while looking through line PN from footpoint P. If the shifted PIL points on the left side are positive (negative) and those on the right side are negative (positive), the filament chirality is sinistral (dextral).

The PIL shift method is used for determining the magnetic field polarities along the left and right sides of the filament axis. Compared to the normal random selection of points or the selection of all points for measuring the polarities on both left and right sides of the filament axis, our method turned out to be more accurate and stable.

\subsubsection{Magnetic Field Helicity}
\label{helicity}
Here we want to compare the automatic detected filament axis chirality with the helicity calculated from the vector magnetograph data in order to discuss the validity of the results. Firstly, the vector magnetograph data are processed to remove the 180$^{\circ}$ ambiguity and the projection effect. The 180$^{\circ}$ ambiguity for the transverse components of the vector magnetic field is resolved by the improved version of the minimum energy method \citep{1994Metcalf,2006Metcalf,2009Leka}. Usually the regions of interest are not close to the center of the solar disk, which leads to the projection effect. We remove this effect with the method proposed by \citet{1990Gary} by projecting those fields from the image plane to the plane tangent to the solar surface at the center of the field of view. Then, we apply the following formula to caculate $\alpha$:
\begin{equation}\label{alpha}
	\alpha = (\partial B_x / \partial y - \partial B_y / \partial x)/B_z.
\end{equation}
Finally, we calculate the average $\alpha$ of the selected region to indicate the sign of magnetic helicity.

\subsection{Detection of the Filament Barb Bearing}
\label{barb}

\subsubsection{Detection of Filament Barbs}
\label{barb detection}
\begin{figure}
\centering
\includegraphics[width=0.4\textwidth,clip=]{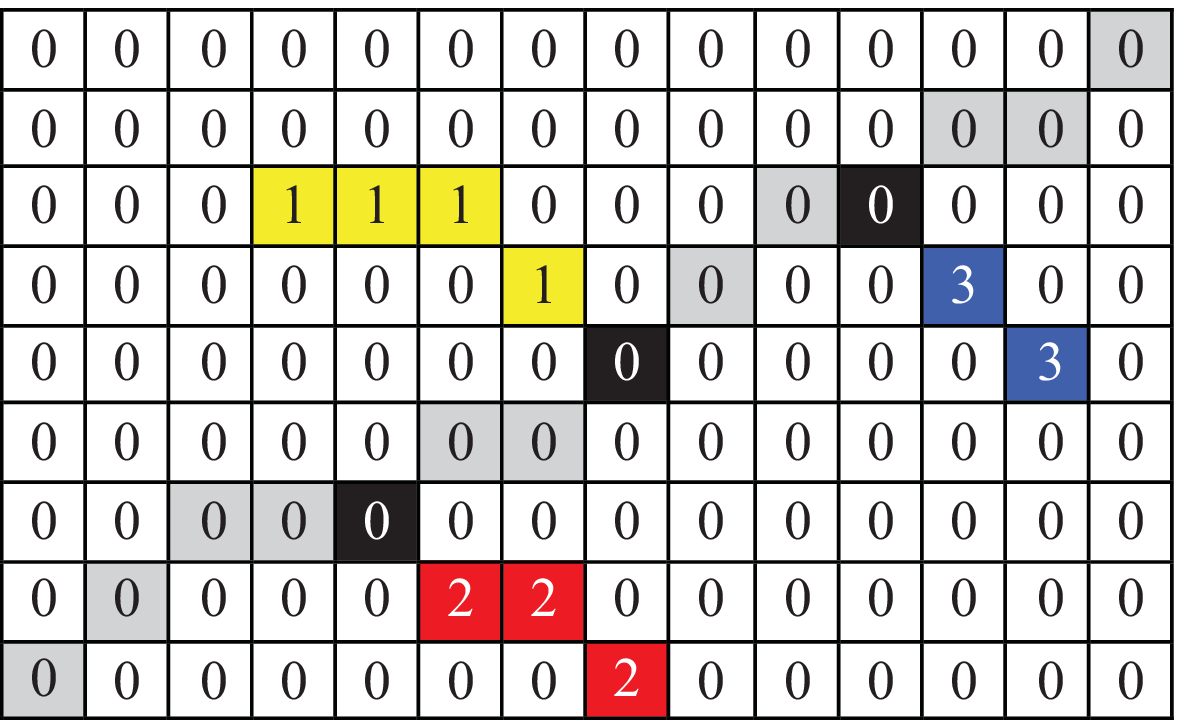}
\caption{An example for explaining the filament barb detection algorithm and the chirality determination method.  The spine node values are changed from 1 to 0, while the grey color is maintained for comparison. The barbs are assigned with unique numbers (1, 2, and 3) and colors (yellow, red, and blue) using the connected components labeling method. The black nodes are the junction nodes between the barbs and the spine that are set as coordinate origins used for the polynomial fitting to determine the barb bearing. All the detailed steps are described in the text.
} \label{fig3}
\end{figure}

Seeing from the binary image of the filament skeleton, we can easily distinguish barbs (branches) from the spine. In other words, there are only barbs in addition to the spine. Through this characteristic, we change all the spine pixels into the background, i.e., setting the spine pixel values to 0 in the binary image of the filament skeleton, where the filament skeleton is the foreground with pixel values being 1. Then, only the separated barbs are left. Using the connected components labeling method, we give each barb a unique number. This method was also used in the automatic detection method for H$\alpha$ full disk filaments and was described extensively in \citet{2013Hao}. The example in Figure~\ref{fig1} (a) is used as before and the detailed process is explained in Figure~\ref{fig3}. We change all the spine nodes to 0. For comparison, the grey color of the spine nodes is maintained. The left nodes are those unconnected barb nodes. We use the connected components labeling method to assign the barbs with unique numbers (1, 2, 3) and colors (yellow, red, blue) as shown in Figure~\ref{fig3} .

%figure 3

\subsubsection{Filament Barb Bearing}
\citet{1998Martin} provided a clear definition of the filament barb bearing. Viewed from either side of a filament, if a barb on the near side veers from the filament spine to the right and downward to the chromosphere, the barb is right-bearing; if a barb veers from the axis to the left and downward to the chromosphere, the barb is left-bearing. From the perspective of the image processing, we can determine the barb bearing by calculating the angle between the spine and the barb. The detailed steps of the method are described as below.

Firstly, we find the junction node between the barb and the spine for each labeled barb. The barb and the spine are in the same connected component, which means that the junction node must be one node in the spine nodes. As shown in Figure~\ref{fig3}, the black nodes are the junction nodes that also belong to the spine. We set the junction node as the coordinate origin. The coordinates of the spine and the associated barb are updated correspondingly.

Secondly, we adopt the polynomial fitting method to fit the barb and the spine, and obtain their inclination angles. Here, we use the first degree polynomial as the default fitting method since the barbs are often short and nearly straight. If a barb is relatively long, the degree of fitting would be changed automatically to a higher one. For example, if  the number of pixels in the barb is more than $20\%$ of that in the spine, the fitting degree would be changed to the second degree; if the number of pixels  is more than $40\%$, the fitting degree would be changed to the third degree, and so forth. Note that unlike barbs, the spine has a curved shape and is much longer than the barbs. We adopt a local fitting method and give a criterion for selecting some local spine nodes for fitting. Namely, we only fit the spine nodes near the junction node instead of the whole spine, which is relatively more accurate. The selection criterion is the number of spine nodes in both directions near the junction node for the polynomial fitting, which is set as the double size of the local barb. Since the junction node (i.e., the coordinate origin) is located in the spine,  it separates the spine nodes close to it into two categories,  which are regarded as left nodes and right nodes for simplicity. In fact, the junction node can be everywhere in the spine except the two footpoints, which means that the number of the left nodes and/or right nodes may be less than the selection criterion. In this condition, the whole left nodes and/or right nodes are selected for the polynomial fitting. Then, we use the slopes of the barb and the spine to calculate the inclination angles. For example, the barb No. 2 of the filament in Figure~\ref{fig3} has 3 nodes. It means that we will choose 6 nodes on each side of the junction nodes, i.e., 6 left nodes and 6 right nodes. Notice that the left side has only 4 nodes that is less than 6, therefore, the whole left side will be selected for the polynomial fitting.

Thirdly, we calculate the angle between barb and spine and determine the chirality by the angle range based on its definition. Using the polar coordinates, the inclination angle of the spine and that of the barb are denoted by $\alpha_s$ and $\alpha_b$, respectively. A barb is recognized to be left-bearing if $\alpha_b-\alpha_s$ is in the ranges of $(-360^{\circ}, -270^{\circ})$, $(-180^{\circ}, -90^{\circ})$, $(0^{\circ}, 90^{\circ})$, and $(180^{\circ}, 270^{\circ})$; it is right-bearing if $\alpha_b-\alpha_s$ is in the ranges of $(-270^{\circ}, -180^{\circ})$, $(-90^{\circ}, 0^{\circ})$, $(90^{\circ}, 180^{\circ})$, and $(270^{\circ}, 360^{\circ})$. Otherwise, it will be marked as an unknown barb bearing. Here, we use the barb No. 1 in Figure~\ref{fig3} as an example to explain the method. The angle between the local barb and the spine is $\alpha_b-\alpha_s\in(90^{\circ},180^{\circ})$, which is in the range for a right-bearing barb. So, the barb No. 1 is determined as a right-bearing barb.

This process goes through all the barbs in a filament, by which we obtain the bearing of each barb. The results are exported to a text file and used for further studies. Note that the shortest-path method can determine the spine with a width of one pixel, but the morphological thinning operation cannot do it so well sometimes. When the spine is changed into the background pixels, the remaining nodes may not be the real barb nodes. They are much smaller than the barb and probably go along the spine. These fake barb nodes will be omitted. Here, we set the length threshold to be 4 pixels. If the number of the barb nodes is less than 4, it is treated as a fake barb and is removed.

\subsection{Data Acquisition and Method Implementation}

Filaments and their barbs are typical features in H$\alpha$ observations. We test the performance of our method with the data from the Big Bear Solar Observatory (BBSO) H$\alpha$ archive (ftp://ftp.bbso.njit.edu/pub/archive). BBSO has an excellent clear sky fraction and high resolution H$\alpha$ full disk image with the size of 2048 $\times$ 2048 pixels and the spatial resolution is $2''$. Thanks to the high spatial resolution, we can identify small-scale barbs clearly. The sample is mainly within solar cycle 24, since the spatial resolution is much higher than before. In order to get the magnetograms in the corresponding time interval,  we use the dataset from the Helioseismic and Magnetic Imager (HMI; \citealt{2012Scherrer,2012Schou}) on board the \textit{Solar Dynamics Observatory} (\textit{SDO}). The \textit{SDO}/HMI has a pixel size of 0.5\arcsec. Since the spatial resolutions of the data are different and the barbs have relative small sizes, the alignment is crucial for further processing. We use our programs to re-identify the solar radius and disk center in pixels in the two data sets. Besides, we manually double check the radius and center location of the Sun in both data to ensure the alignment. The error of the co-alignment is less than $1''$.

Our detection method for the filament barb bearings is developed by using the MATLAB Desktop Tools and Development. In addition to using MATLAB to implement the algorithm for the filament axis chirality, we also use IDL with Solar SoftWare (SSW) library to pre-process the \textit{SDO}/HMI magnetograms. The time efficiency of these methods are millisecond level for the whole processing including the pre-processing of an H$\alpha$ full disk filtergram.

%figure 4
\begin{figure*}
\centering
\includegraphics[width=0.9\textwidth,clip=]{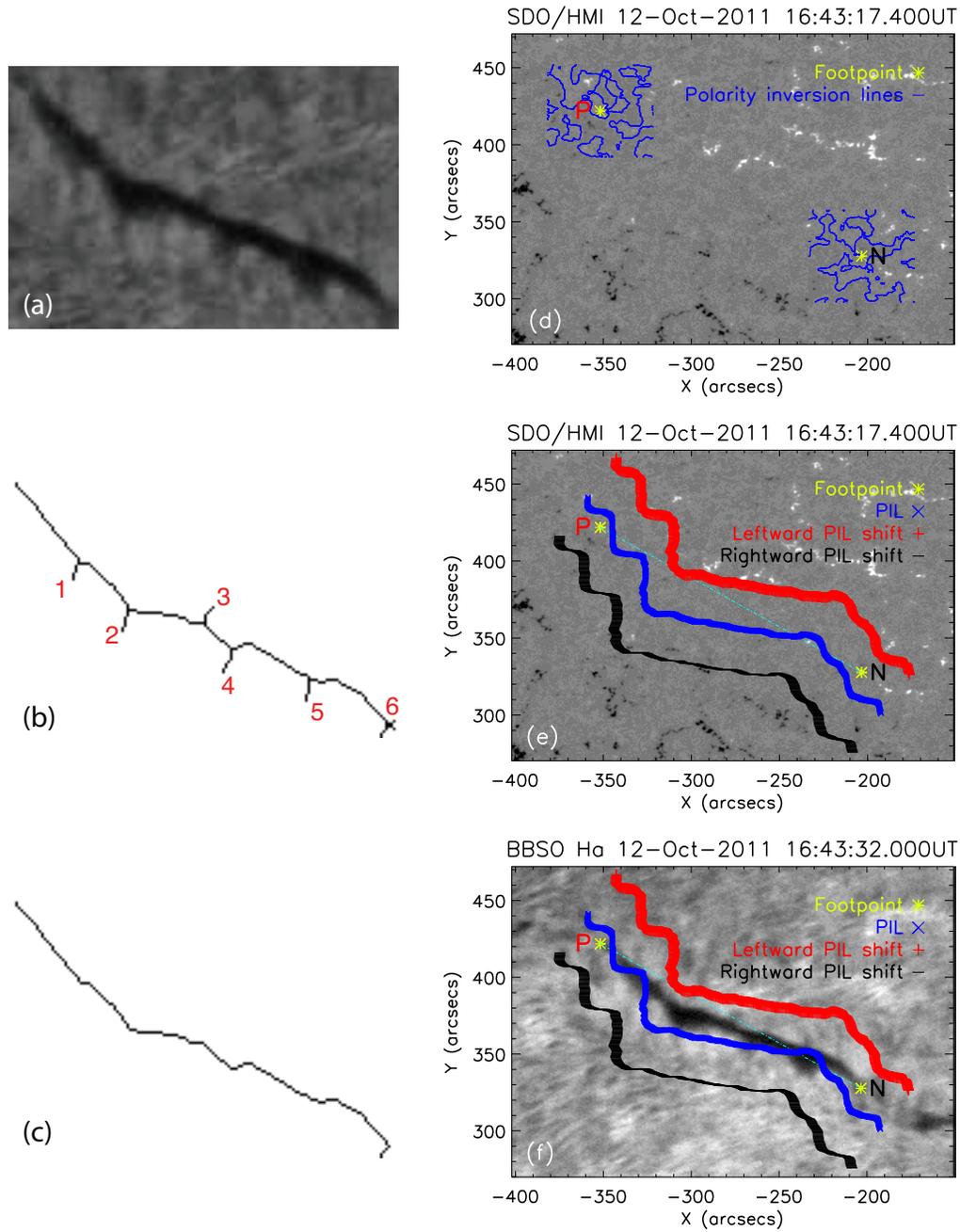}
\caption{A filament observed at 16:43:32 UT on 2011 October 12, located in the northern hemisphere. (a) The extracted enhanced H$\alpha$ image. (b) The skeleton of the filament, each barb is labeled with a unique number. (c) The spine of the filament. (d) The  \textit{SDO}/HMI magnetogram where the filament is located in. The yellow asterisks show the footpoints of the filament. The two footpoints are designated by P and N, respectively, meaning positive and negative polarities. The blue line shows the contour of the polarity inversion lines (PILs) around the footpoints. (e) The yellow asterisks show the footpoints of the filament as panel(d). The blue multiplication signs show the PIL. The red plus signs show the positive shift of the PIL and the black minus signs show the negative shift of the PIL. The shift direction is perpendicularly to the line (light blue dash-dotted line) that connected two footpoints of the filament. (f) The H$\alpha$ filament image overlaid by PIL and its shift lines. The meaning of the signs are the same as that in panel (e).
} \label{fig4}
\end{figure*}

%figure 5
\begin{figure*}
\centering
\includegraphics[width=0.9\textwidth,clip=]{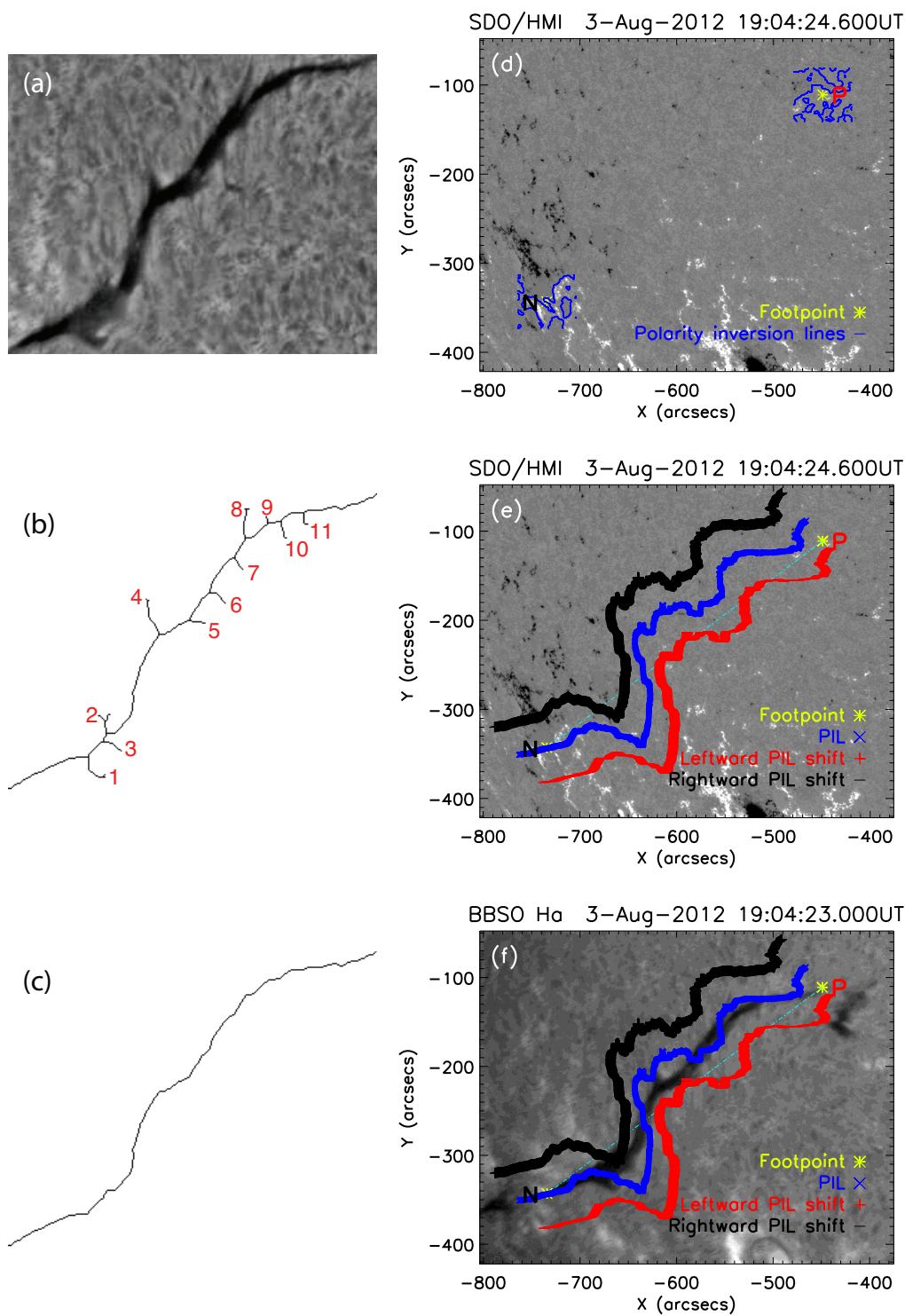}
\caption{Similar to Figure~\ref{fig4} but for the filament observed at 19:04:23 UT on 2012 August 3 and located in the southern hemisphere.
} \label{fig5}
\end{figure*}

%figure 6
\begin{figure*}
\centering
\includegraphics[width=0.9\textwidth,clip=]{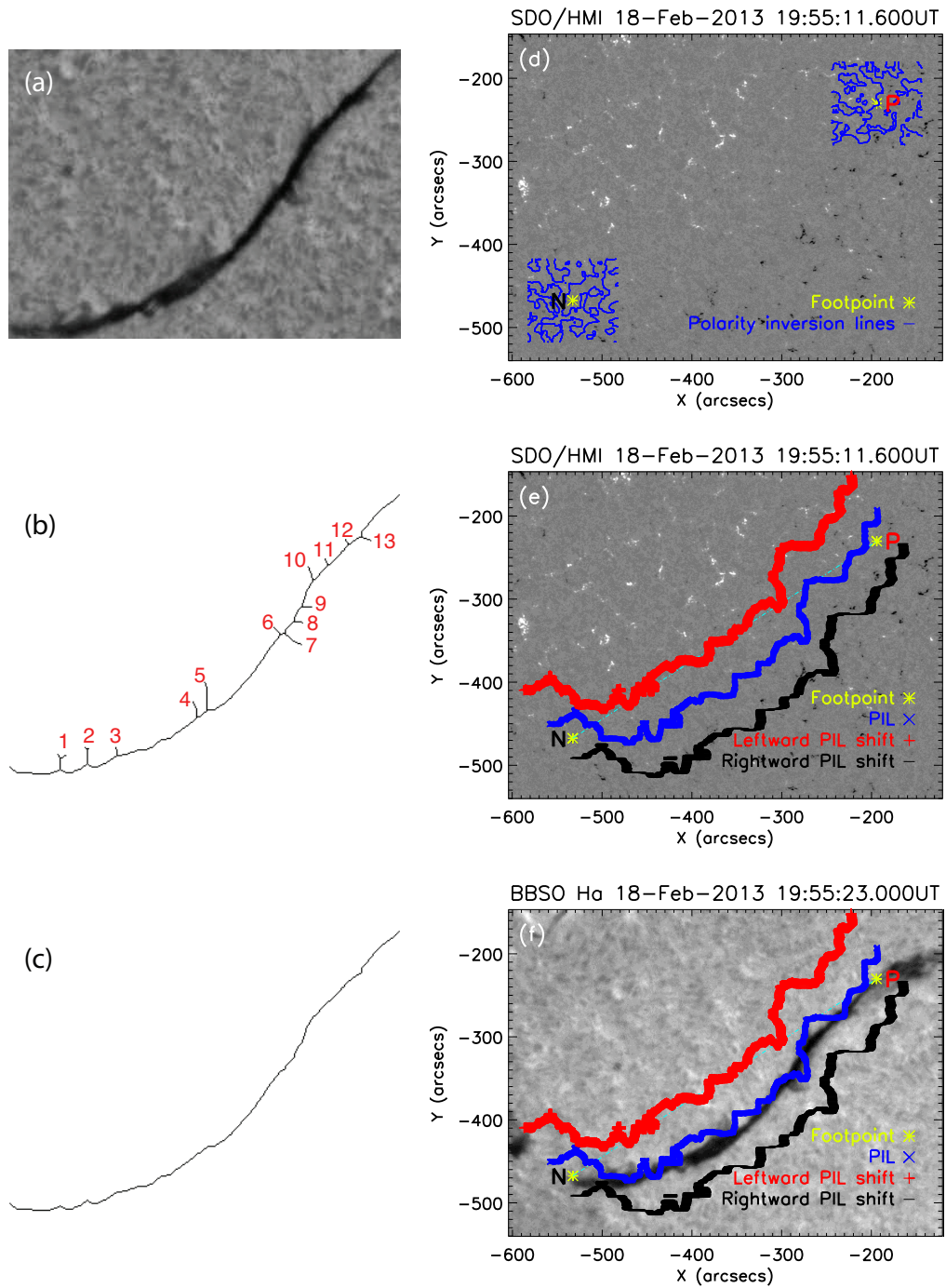}
\caption{Similar to Figures~\ref{fig4} and \ref{fig5} but for the filament observed at 19:55:23 UT on 2013 February 18 and located in the southern hemisphere.
} \label{fig6}
\end{figure*}

%figure 7
\begin{figure*}
\centering
\includegraphics[width=0.9\textwidth,clip=]{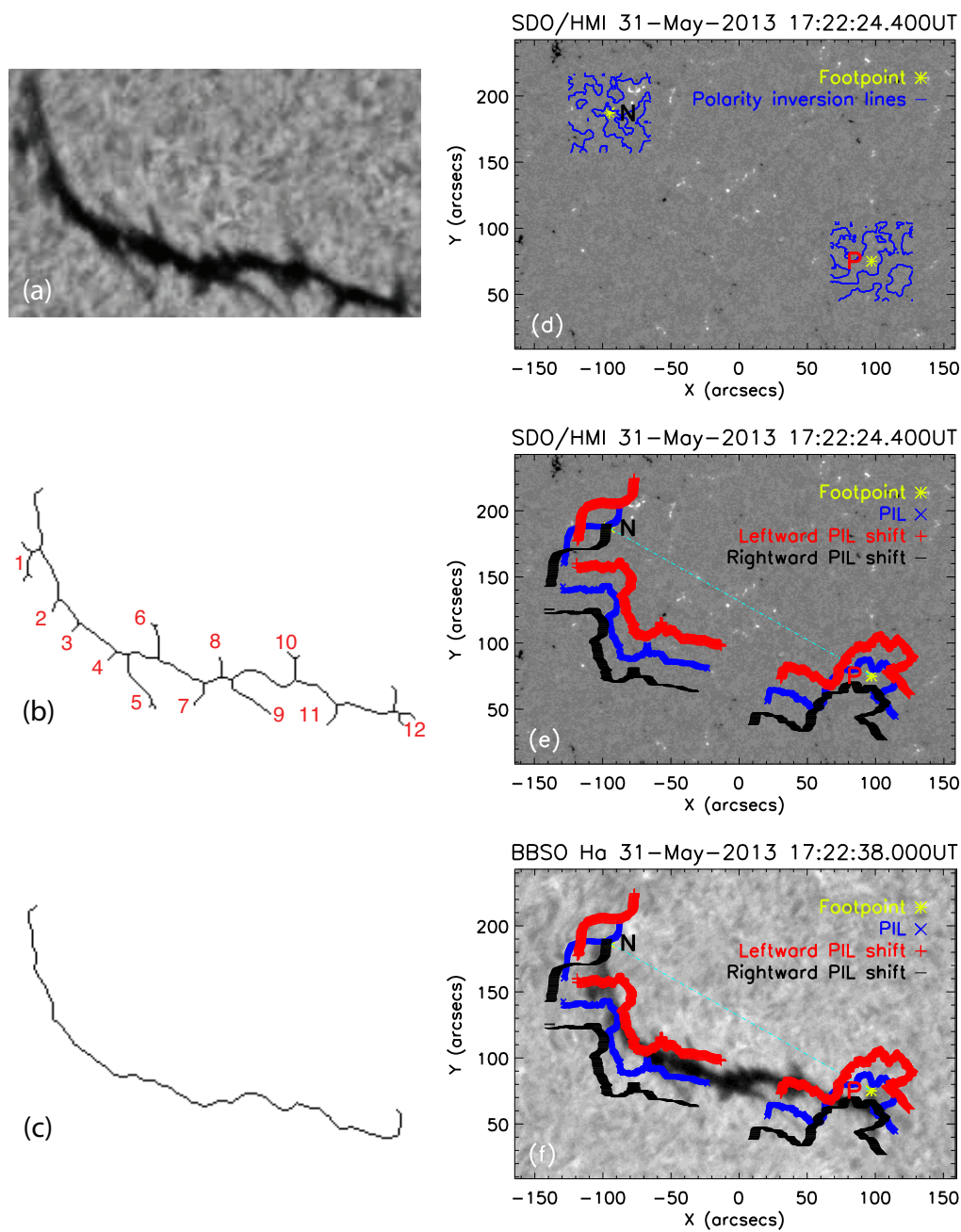}
\caption{Similar to Figures~\ref{fig4} and \ref{fig5} but for the filament observed at 17:22:38 UT on 2013 May 31 and located in the northern hemisphere.
} \label{fig7}
\end{figure*}

\begin{table*}
\centering
\caption{Automatic detection results of the filament chirality and barb bearing for the examples.} \label{table}
\begin{tabular}{lclclclclclclc} \\
\hline \hline
 \noalign{\smallskip}
Date & Time   & Hemisphere  & Average $\alpha$ & Ref. number of  & Ref. number of & Filament chirality \\
     &	  (UT)      &	      &    & Left-bearing barbs & Right-bearing barbs & determined by the wire model  \\
 \noalign{\smallskip}
\hline
 \noalign{\smallskip}
2011 Oct 12   & 16:43:32 & Northern  & $-$    &             &  1, 2, 3, 4, 5,  & Sinistral \\
		      &	                     &		     &          &  	      & 6                    & 	             \\
 \noalign{\smallskip}
\hline
 \noalign{\smallskip}
 2012 Aug 03   & 19:04:23   & Southern  & $+$    & 2, 4, 6, 7, 8,     &1, 3, 5      & Sinistral \\
		       &	                &		 &          &  9, 10, 11	      &                & 	       \\
 \noalign{\smallskip}
\hline
 \noalign{\smallskip}
2013 Feb 18   & 19:55:23   & Southern  & $+$      & 1, 2, 3, 4, 5,    & 7, 8, 9, 13          & Dextral  \\
 		       &		       &		&	       & 6, 10, 11, 12    &                          &  \\
 \noalign{\smallskip}
\hline
 \noalign{\smallskip}
 2013 May 31   &17:22:38    & Northern  & $-$      & 4, 7, 11           & 1, 2, 3, 5, 6,   & Dextral \\ 		
		     &	                         &		 &           &    		      & 8, 9, 10, 12    &             \\
% 			&		             &		     &	                      &             &                         &  C  \\
 \noalign{\smallskip}
\hline
\end{tabular}

\end{table*}

\section{Results}
\label{results}

We select four example filaments observed at 16:43:17 UT on 2011 October 12, at 19:04:24 UT on 2012 August 3, at 19:55:11 UT on 2013 February, and 17:22:24 UT on 2013 May 31, as shown in Figures~\ref{fig4}--\ref{fig7}. Two of them are located  in the northern hemisphere, and the other two are in the southern hemisphere. They are all quiescent filaments, which have relatively large-scale and stable structures. Therefore, they can be observed with relatively  high resolution for studying the fine structures. Since these four filaments are processed by the same process steps, here we only take the filament observed on 2011 October 12 in Figure~\ref{fig4}  as an example to explain the processed results for convenience. After pre-processing the H$\alpha$ full disk images, we obtain a sub-image of a filament as shown in Figure~\ref{fig4}(a). The binary image of its skeleton is shown in Figure~\ref{fig4}(b).  Next, our spine detection method is used to find the filament spine as shown in Figure~\ref{fig4}(c). We can see the elongated spine as thin as one pixel and without any barb pixels. Then we change the view to the co-aligned magnetogram in Figures~\ref{fig4}(d) and \ref{fig4}(e). The two footpoints are shown as yellow asterisks and the PIL contours around the footpoints are shown as blue lines in Figure~\ref{fig4}(d). The PIL coordinates along the filament spine are calculated and obtained, which are shown as the blue cross signs in Figures~\ref{fig4}(e) and \ref{fig4}(f). The program shifts the PIL coordinates along the direction perpendicular to the straight line connecting the two footpoints, and  the polarities of the shifted PILs on both sides are measured, as shown in Figures~\ref{fig4}(e) and \ref{fig4}(f). The red plus signs show the shifted PIL on the positive polarity and the black minus signs show the shifted PIL on the negative polarity. In the meantime, the program measures the corresponding magnetic polarities of the two filament footpoints,  which are derived by the automatic spine detection method from the H$\alpha$ filtergram, as shown by yellow asterisks in Figures~\ref{fig4}(d), \ref{fig4}(e) and \ref{fig4}(f). Finally the filament axis chirality is determined by the method proposed by \citet{1994Martin}. For the filament barbs, the program changes all the spine pixels into the background ones and uses the connected components labeling method to give each barb a unique number as shown in Figure~\ref{fig4}(b). This filament has 6 barbs in total. Each barb is labeled with a unique number for reference. Then, a polynomial fitting method is adopted to fit each barb and the local part of the spine, and we obtain the angle between them. The barb bearing is determined by the angle  based on its definition. After the whole process for the filament is completed, the position coordinates, the length, and the chirality of each barb are recorded in a text file. 

We also analyze the vector magnetic fields. The selected time and the field of view for the analysis are the same as Figure~\ref{fig4}(d). All results of the four examples are summarized in Table~\ref{table}. Since the filament is located in the quiescent region, the magnetic field is relatively weak compared to that in active regions. Therefore, we test the reliability for the average $\alpha$ computation by a Monte Carlo method. The measurement uncertainties of the vector magnetic field have been provided by the HMI vector magnetic field pipeline \citep{2014Hoeksema,2014Centeno}. For each component (strength, inclination, and azimuth) of the vector magnetic field, we generate normally distributed random noises, whose standard deviation is 1. Next, the random noises are multiplied by the measurement uncertainties of the vector magnetic field and added to the original value of the vector field. We remove the 180$^\circ$ ambiguity and correct the projection effect for the noise added vector magnetic field. Then, the force-free parameter $\alpha$ is computed for the region of interest. We repeat the above processes for ten times and find that, for all the ten cases, the force-free parameter $\alpha$ is negative for the 2011 October 12 event. Similarly, we analyze the other three events. The finally computed average $\alpha$ is the same as that listed in Table 1. The uncertainty analysis shows that the derived average force-free parameters for the quiescent filaments are reliable considering the measurement uncertainties of the vector magnetic field.

The filaments observed on 2011 October 12 and on 2013 May 31 are located in the northern hemisphere, both of them have negative average $\alpha$, while the detected axis chiralities based on the wire model are opposite. The filaments observed on 2012 August 3 and on 2013 February 18 are located in the southern hemisphere, both of them have positive average $\alpha$, while the detected chiralities based on the wire model are opposite, too. The four examples show that the average $\alpha$ is negative in northern hemisphere and positive in southern hemisphere. It indicates that the helicity in the northern hemisphere is negative and that in the southern hemisphere is positive. From the automatic detected barb bearings in Table~\ref{table},  we can find that the filaments located in the northern hemisphere mainly have right-bearing barbs and the filaments located in the southern hemisphere mainly have left-bearing barbs.  Note that except the erroneously detected barbs, often there are both left-bearing and right-bearing barbs in a single filament. It is most obvious in Figure~\ref{fig5}. Compared with the original H$\alpha$ filtergraph in Figure~\ref{fig5}(a) with the automatic detected barbs in Figure~\ref{fig5}(b), the automatic results show that the barbs Nos. 2, 4, and 8 are left-bearing barbs,  whereas barbs Nos. 1 and 5 are right-bearing barbs, which we can also directly distinguish manually based on the definition by \citet{1994Martin}.

From the results of these four example filaments and other ones we tested, we find that the accuracy of our automated detection method for the bearing of filament barbs is around $80\%$. The false bearing detections are largely due to the following two reasons. Firstly, the detected angle between the barb and the spine is nearly 90$^{\circ}$, which is hard to tell the chirality automatically, even one cannot determine the chirality by human eyes sometimes. In this case, our program still outputs the automatically detected results. However, it gives a remark to such a kind of barbs, indicating that the barb bearing needs to be further verified. Table~\ref{table} shows the results determined only by the automatic program without further manual recognition. Secondly, the barbs sometimes have branches, while the program just regards them as a whole structure for the polynomial fitting, so the detected result is an average. For example, the barb No. 1 shown in Figure~\ref{fig7}(b) has a ``Y"-shaped structure. Our program has the ability to process the barb's branches separately if the data have a very high resolution as well. Since these conditions appear only a few times in the current database, here we still process the barb as a whole one. The chirality of the filaments can also be detected by our automatic program. However, it is found that the sign of the chirality determined with the method proposed by \citet{1994Martin} is sometimes contradictory to that calculated from vector magnetogram.

\section{Discussions and Conclusion}
\label{conclusion}

To our knowledge, our method is the first one that combines H$\alpha$ filtergrams and magnetograms to automatically detect and determine the filament chirality and the barb bearing among the existing filament chirality automatic detection methods. This method can determine the spine and the barbs without changing the filament topological structure. In other words, it does not introduce any other features or reduce information from the filament skeleton. The only crucial points are the filament skeleton obtained by the previous processing and the alignment between the H$\alpha$ filtergram and the magnetogram.  If the skeleton is not as thin as possible, the angle between the barb and the spine may be calculated mistakenly, which further influences the determination of the barb bearing.

\citet{2003Pevtsov} studied 2310 filaments observed at BBSO during 2000--2001 and showed that a single filament may have both left-bearing and right-bearing barbs. We got similar results. In their study, they noticed that apart from the filaments with pure bearing (i.e., all barbs within a filament have the same bearing) occupying $79\%$ of all their analyzed filaments, there are $21\%$ of all filaments with mixed bearings (both left-bearing and right-bearing barbs exist within the same filament).  It implies that this kind of filaments are not as prevalent as the filaments with pure bearing, but they do exist. It is argued by \citet{2008Martin} that a barb with opposite bearing with other barbs in one filament may be an illusion owing to the low spatial resolution of the observation. With the high-resolution observation by the Swedish Solar Telescope (SST), they illustrated that a barb, which is right bearing in the low-resolution BBSO images, actually consists of many left-bearing threads. Of course, there is such a possibility. However, more frequently the barbs with the bearing opposite to other barbs in one filament are real, since part of the filament channel can be a flux rope, with the remaining part being a sheared arcade, and the barbs associated with these two parts have opposite bearings despite of the same sign of helicity \citep{2010Guo1,2014Chen}.

The fact that a non-negligible fraction of the filaments has barbs with mixed bearings provides a further piece of evidence against the notion that there is one-to-one correspondence between filament chirality and the barb bearing. According to the wire model, a filament with mixed barb bearings has opposite helicity along the filament spine. However, it is often believed that the whole filament and the filament channel have the same sign of helicity \citep{1994Rust2,1995Low,1996Low,1999Rust,2001Martens}. Regarding the mixing bearings of barbs in one filament, \citet{2010Guo1} examined the extrapolated nonlinear force-free magnetic field in the corona and found that the filament has both flux rope and sheared arcade configurations \citep[see][as well]{2002Aulanier}. The barbs of the filament, which has negative helicity, are left-bearing at the sheared arcade parts, and are right-bearing at the flux rope parts. \citet{2014Chen} proposed a unified paradigm, i.e., a sinistral filament can have both left-bearing barbs when the hosting field lines have an inverse-polarity magnetic configuration and right-bearing barbs when the hosting field lines have a normal polarity configuration; A dextral filament can have both right-bearing barbs when the hosting field lines have an inverse-polarity magnetic configuration and left-bearing barbs when the hosting field lines have a normal polarity configuration. The existence of barbs with opposite bearings in one filament implies that some of the quiescent filaments have a complex magnetic structure that are formed partly by a flux rope and partly by a dipped arcade. One cannot determine the whole quiescent filament chirality only by the bearings of the barbs. The detection of filament chirality should combine both morphology and magnetic field observations.

The calculated average $\alpha$ shows that the helicity is negative in the northern hemisphere and positive in the southern hemisphere. It means that the helicity signs given by the average $\alpha$ of the four example filaments are in agreement with the so-called ``hemispheric helicity pattern", i.e., solar magnetic fields generally have positive helicity (right-handed twist of the fields) in the southern hemisphere and negative helicity (left-handed twist) in the northern hemisphere. Compared with the automatic detection results, the barbs of the two filaments in the northern hemisphere are mainly right-bearing and those of the other two in the southern hemisphere are mainly left-bearing. According to the paradigm in \citet{2014Chen}, it implies that the magnetic structures of these filaments are mainly flux ropes, which is consistent with the results on the filament magnetic structure obtained by previous authors \citep{1994Rust2,1995Low,1998Aulanier}.

We agree that the definitions of the filament chirality and barb bearing by \citet{1994Martin} and \citet{1998Martin} are reasonable and practical. We strictly follow their definitions to determine the chirality of the filament and the bearing of the barbs. However, we found that the helicity of a filament determined by magnetic polarities of the two footpoints of the filament, as proposed by \citet{1994Martin}, may be the same as or opposite to that calculated from the vector magnetograms. There are several reasons for it. First, there might be no solid physical basis to claim that the footpoints of a filaments are the intersection points of the magnetic field lines with the solar surface. Especially when the filament is of inverse polarity, the footpoints of an H$\alpha$ filament are located on the opposite side of the PIL compared to the footpoints of the magnetic field lines, as illustrated by panels (a) and (b) in Figure 7 of \citet{2014Chen}. Second, the footpoints of some filaments are too weak to be detected accurately. Third, when the barbs are near the end of the spine, it is hard to distinguish whether it is the end of the spine or the barbs near the junction node. As the right end of the filament in Figure~\ref{fig7}(a), the shortest path method determines it as a part of the spine, while empirically,  we would recognize it as a barb. These cases are only found in the ends of the filament when the end junction node has more than one branch. Therefore, it does not affect the barb detection within the filament skeleton. In such a case, one way to determine which branch belonging to the spine is checking the curvature: if the branch connected with the spine have the smallest curvature, this branch and the spine will be determined as the whole spine, and the others will be determined as barbs. We will validate the method and improve our spine detection method in our future version. Hence, the method by measuring the magnetic filed of the two footpoints to determine the filament chirality (i.e., the helicity), as proposed by \citet{1994Martin}, may get wrong results without the help of vector magnetogram data. In the case that vector magnetograms are not available, it would be better to determine the helicity by the skewness of coronal loops or EUV dimmings \citep{2011Jiang}. Even when all these alternative choices are not available, we can still assume the flux rope magnetic field model as a tradeoff since a majority of filaments are of inverse-polarity type. Of course, the results of \citet{2010Guo1} and \citet{2014Chen} imply that this approximation method may lead to errors sometimes. Therefore, a large amount of data are still needed to study the statistical results to validate these methods.

In summary, we propose an efficient method for automatically detecting the bearing of solar filament barbs. This method employs the unweighted undirected graph concept and the Dijkstra shortest-path algorithm to recognize the filament spine, and the connected components labeling method to identify the barbs. The results of the relatively high resolution data verified that the performance of our method is accurate and robust. Since our method is based on our previous automatic detection method for H$\alpha$ filaments in full-disk images, it can process not only a given filament in the partial frame, but also multiple filaments in the full disk images. Four quiescent filaments are automatically detected and illustrated to show the results. A filament with a unique chirality may have different barb bearings in its different parts. It infers that some quiescent filaments have a complex magnetic structure that is formed partly by a flux rope and partly by a dipped arcade. One could not determine the filament chirality and magnetic helicity only by the barb bearing. The determination of the filament chirality can be done by combining both morphological and magnetic field observations.  In the condition of lacking vector magnetic field data, the method of determining the filament axis chirality based on the magnetic field signs of the two footpoints is not valid. The detection of the footpoints are subject to the magnetic configuration and filament material distribution, which means that the footpoints are hard to exactly locate. The best way to judge the filament chirality is to examine the skewness of the flare loops after the filament eruption (e.g., \citealt{2011Jiang}) or the skewness of the mass drainage as proposed by \citet{2014Chen}. In the future, we will focus on these methods to automatically detect the filament chirality based on the H$\alpha$ observations by the Optical and Near-infrared Solar Eruption Tracer (ONSET, \citealt{2013Fang}).

\normalem
\begin{acknowledgements}
We thank Big Bear Solar Observatory (BBSO) team and New Jersey Institute of Technology (NJIT) for making the data available and the magnetograph data are courtesy of SDO and the HMI science teams. This work was supported by NKBRSF under grants 2011CB811402 and 2014CB744203, NSFC grants 11203014 and 11025314, as well as the grant from CSC201306190046  and CXZZ130041.

\end{acknowledgements}

%\bibliographystyle{raa}
%\bibliography{bibtex}

\end{document}